\documentclass[12pt, final]{iopart}
\usepackage[]{epsf}
\usepackage[]{graphics}

\def\vec#1{\mathchoice{\mbox{\boldmath$\displaystyle\bf#1$}}
{\mbox{\boldmath$\textstyle\bf#1$}}
{\mbox{\boldmath$\scriptstyle\bf#1$}}
{\mbox{\boldmath$\scriptscriptstyle\bf#1$}}}

\begin{document}
\title{Portfolio Selection with Probabilistic Utility, Bayesian
Statistics, and Markov Chain Monte Carlo}
\author{Pietro Rossi\S\footnote[1]{From Jan 1$^{st}$, 2003 Prometeia S.r.l., Via Marconi 43, 
40100 Bologna, Italy.},
 Massimo Tavoni\ddag, Flavio Cocco\ddag\, and Robert Marschinski\P}
\address{\S\ ENEA-HPCN, Via Martiri di Monte Sole 4, 40100 Bologna, Italy}
\address{\ddag\ Prometeia S.r.l, Via Marconi 43, 40100 Bologna, Italy}
\address{\P\ Institute for Physics, University of Potsdam, Potsdam, Germany}

\ead{pietro.rossi@prometeia.it, rossi@bologna.enea.it,
     massimo.tavoni@prometeia.it,
     flavio.cocco@prometeia.it,
     robert.marschinski@pik-potsdam.de}

\begin{abstract}

We propose a novel portfolio selection approach that manages to ease some of the problems 
that characterise standard expected utility maximisation.
The optimal portfolio is no longer defined as
the extremum of a suitably chosen utility function: the latter, instead, is reinterpreted as
the logarithm of a probability distribution for optimal portfolios and the selected
portfolio is defined as the expected value with respect to this distribution.
A further theoretical aspect is the adoption of a Bayesian inference framework.
We find that this approach has several attractive features, when comparing it
to the standard maximisation of expected utility.
We remove the over-pronounced sensitivity on external
parameters that plague optimisation procedures and obtain a natural and 
self consistent way to account for uncertainty in knowledge and for personal views.
We test the proposed method against traditional expected utility maximisation, using artificial 
data to simulate finite-sample behaviour, and find superior performance of our procedure.
All numerical integrals are carried out by using Markov Chain Monte Carlo,
where the chains are generated by an adapted version of Hybrid Monte Carlo.
We present numerical results for a portfolio of eight assets using historical 
time series running from January 1988 to January 2002.

\end{abstract}
\maketitle

\newpage

\section*{Introduction}

Classical portfolio selection \cite{Markovitz} by \underline{M}aximisation of 
\underline{E}xpected \underline{U}tility (MEU) suffers from 
well-documented drawbacks \cite{Michaud}:
it often leads to extreme and hardly plausible portfolio weights, which additionally are very sensitive 
to changes in the expected returns.  
Moreover, it does not take into account differences in the level of uncertainty 
associated with the various input variables (estimation-errors), since its straightforward optimisation 
procedure imposes infinite faith on the estimated parameters.
Historical data provides some information on future returns, but it is well known that simple-minded 
use of this information often leads 
to nonsense because estimation disturbance overwhelms the value of the information contained in the data. 
In fact, the positions of extrema of a function are often highly sensitive to irrelevant distribution details
and it is thus quite simple to build examples (see following section) where a minimal parameter variation induces a very large shift
in the extrema location.

The issue of uncertainty in expected returns and its implications for portfolio selection has been extensively analysed 
in the relevant literature: starting with the work of Bawa, Brown and Klein \cite{Bawa}, many authors have since addressed
the problem, often resorting to a Bayesian framework \cite{Barry, BlackLitterman, Jobson, Jorion, Frost, Michaud, Ziemba}. 
More recently, with
a growing debate on asset return predictability  (which will not be addressed here), 
the issue has re-gained the attention of the academia \cite{Balduzzi, Barberis, Brennan, Johannes, TerHorst}.

Nevertheless, parameters determined by observation of historical data are not the only source of trouble for portfolios
based on function optimisation: all the expected utility maximisation procedures suffer from the presence of a scalar
parameter related to the investor's risk aversion,  whose value cannot be set by the theory but still 
sensitively affects the resulting portfolio composition. Actually, due to a complete lack of scale for this 
risk-aversion parameter, it is usually adjusted ex post by hand, i.e. by merely observing where ``the dynamics happen''
and defining an ad hoc scale according to the simple prescription ``increase the parameter if you want a more aggressive - 
meaning riskier - portfolio''. In some cases this might be
an acceptable ``degree of freedom'', allowing to customise portfolios, but when combined with the very parameter sensitive
maximise-expected-utility-optimisation (MEU in the following), it turns out to produce highly unstable 
and inconsistent portfolios, meaning
that a portfolio might change significantly for an apparently small shift in risk-aversion, and might even be
less ``aggressive'' than a neighbouring portfolio with a lower risk-aversion. This will be discussed in more detail 
in Section \ref{results}. To our knowledge, this relation between optimisation procedure 
and risk-aversion parameter has not been investigated in previous studies.

The primary objective of this paper is to offer a common prescription for easing 
both of these pathologies. 
In order to eliminate the intrinsic optimisation instability caused by the over-sensitiveness towards external parameters, 
we suggest a different interpretation 
of the utility function. We consider the utility function to be the logarithm of the probability density for the portfolio to assume 
a given composition, and we define as optimal the expected value of the portfolio's weights with respect 
to that probability. As will be shown, this leads to an improved, more robust portfolio selection procedure, 
which allows us to incorporate the risk-aversion parameter in a stable and - even if not theory determined - at least self-consistent manner. 

As for the issue of uncertainty in parameter determination, we adopt a fully Bayesian 
approach, in which parameters characterising the distribution of the data are described by distributions themselves. 
Additionally, the Bayesian approach offers a natural framework for the incorporation of subjective investor views
into the portfolio selection procedure. 
Finally, through this method uncertainty is taken into account by stating explicitly the errors associated with the 
determination of the portfolio.

In what follows, we first introduce and discuss a theoretical framework in general terms. 
When coming to the specification of the posterior distribution and of the utility function, we resort to a 
multivariate Gaussian distribution framework, in line with common practice, deferring relaxation of this 
assumption to future research.

The final contribution of our paper concerns the numerical technology employed to perform all the relevant integrals.
Most of these cannot be computed explicitly and therefore we will resort to a dynamical Monte Carlo integration 
or ``Markov Chain Monte Carlo''. To enhance performance we have used a variation of the Metropolis-Hastings prescription
known as ``Hybrid Monte Carlo", that first appeared in the physics literature in 1987 \cite{Duane87}. A brief outline of the
algorithm is sketched in Section \ref{num} and we refer the reader to the appendix for a more extensive discussion.

For testing the performance of our proposed method, we use artificial data derived from known multivariate Gaussian distributions,
calibrated using data from eight different asset classes for the last 14 years.
This allows to simulate the finite-sample behaviour of our ``best portfolio" estimator (PU from now on), and compare it to the standard MEU
prescription. Since the real optimal portfolio - with respect to the chosen utility function - is known, the speed of 
convergence can be measured empirically. As will be shown in Section \ref{results}, our method clearly outperforms the 
simplistic optimisation.
Interestingly one observes that up to a 
threshold of about 350 monthly observations (corresponding to almost 30 years of data)
the knowledge gained from data is actually insufficient for selecting any but the uniformly distributed portfolio. We are also able to 
confirm a significant
improvement with regard to the instability of the algorithm induced by the risk-aversion parameter.
In the end some backtesting is performed: when
looking at ``what if'' investment scenarios, our method again shows superior performance for at least one typical investment profile.

The paper is organised as follows: in Section \ref{meth} we propose our method, in which MEU optimisation is replaced 
by a double expectation with respect to (a transformation of) the utility function and the conditional posterior distribution.
Section \ref{post} is devoted to the analysis of the posterior distribution. In Section \ref{num} 
we deal with numerical integration, and report empirical results in Section \ref{results}. 
Conclusions and final remarks are presented in Section \ref{concl}.

\section{The ``Recommended'' Portfolio Approach}
\label{meth}
After illustrating some typical features of the problem under examination by means of a simplified example, 
we will formally introduce our probabilistic interpretation of the utility function, followed then by the Bayesian analysis
of the problem. 

\subsection*{A simplified example}
\begin{figure}[h]
\begin{center}\scalebox{.6}{ \epsfbox{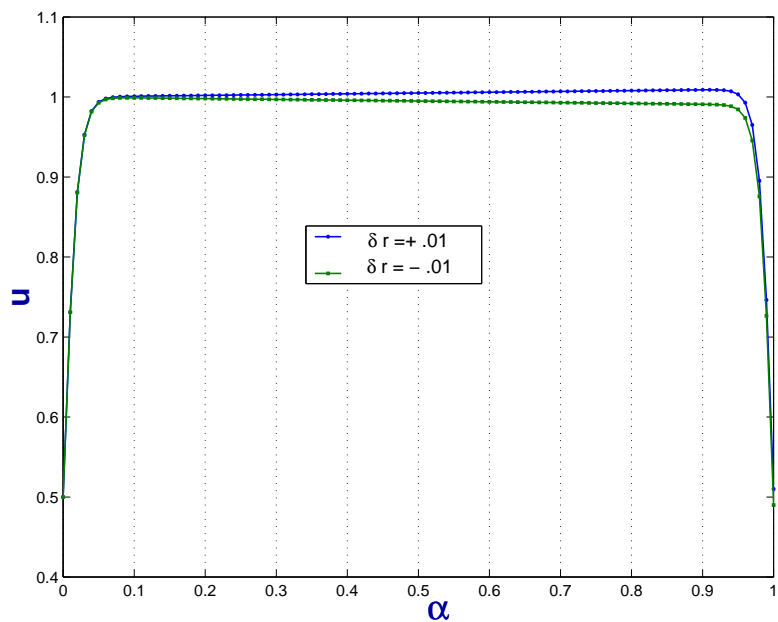} }
 \caption{\label{plot} Plot of the function $u(\alpha) $ from Eq.\ref{fsample}.}
\end{center}
\end{figure}

To illustrate what we said in the introduction, let us consider the following function:
\begin{equation}
\fl
u(\alpha, M, \delta r) = (1+ \alpha \delta r)\bigg[1+\exp\bigg(-\frac{\alpha}{M}\bigg)\bigg]^{-1}
   - \alpha^2\bigg[1+\exp\bigg(-\frac{\alpha-1}{M}\bigg)\bigg]^{-1}
\label{fsample}
\end{equation}
which is plotted twice in its dependence on $\alpha$ in Fig.\ref{plot}. Both graphs correspond to a value of $M=.01$;
they only differ in the choice of $\delta r$ that is set to .01 and to -.01.

With these minimal parameter modifications, the qualitative features of the two curves are
virtually unchanged, but the maximum within the interval $\alpha \in [0,1]$ moves from $\alpha_{max} = 0.91$ 
to $\alpha_{max}= 0.09$.
Clearly this represents a serious
problem whenever the determination of parameters would be based on a fit to observed
data which necessarily will be plagued by errors. The value for $\delta r$ could, for example,
be the expected excess return of our portfolio with respect to some low risk asset: a small error in $\delta r$ would have a
severe effect on the location of the maximum and, consequently, on the selected value for $\alpha_{max}$.
If the function $u(\alpha)$ were to represent some sort
of expected utility, it should be pointed out that just choosing a portfolio (mimicked here by the choice
of a value for $\alpha$) by blind faith according to the maximising principle, would, for both values of $\delta r$, 
lead to an unnecessary high amount of risk, since
a minimal error in $\delta r$ could cause a deep fall
respectively to the left or right of the two maxima. For both functions a more conservative settlement
	somewhere in the middle would not induce nearly as much risk and still achieve results not too different from those
	guaranteed by the apparent ``optimal'' choice.  

In order to overcome this problem we suggest to interpret the expected utility as proportional to the
logarithm of a probability in the space of portfolios, and replace the prescription of selecting the ``optimum''
portfolio as the maximum of the expected utility by the rule that the recommended portfolio
is the expectation value of the portfolio weighted by its probability:

\begin{equation}
\label{xavg}
	E(\alpha|M, \delta r, \nu)  =  Z(M,\delta r,\nu) \int_0^1 d\alpha \quad\alpha\quad \exp\bigg( \nu u(\alpha, M, \delta r) \bigg),
\end{equation}
where
\[
 Z^{-1}(M, \delta r, \nu) \equiv \int_0^1  d\alpha \exp\bigg( \nu u(\alpha, M, \delta r) \bigg),
\]
and $\nu$ is a constant. This definition of $E(\alpha|M, \delta r, \nu)$ is a continuos and differentiable
function in all of its parameters as opposed to the discontinuity of the maximum prescription, and if we define
\[
	\alpha^* = \lim_{\nu \to \infty} E(\alpha| M, \delta r, \nu),
\]
it is easy to show that $\alpha^*$ is the solution of
\[
	\mbox{MAX}_{\alpha} \quad u(\alpha, M, \delta r),
\]
and we hence correctly recover the singular behaviour of maximisation as 
the limit value for perfectly analytic functions. 
The reader reluctant of replacing the
maximisation principle can always interpret our prescription as a smooth interpolation 
procedure, much in the same spirit as the simulated annealing minimisation procedure introduced by 
Kirkpatrick \etal\cite{Kirkpatrick83}.
In the remaining parts of this paper we will sometimes abuse the term ``optimum''
referring with it also to the portfolios produced by our approach. From the context it will
always be clear what kind of optimum, MEU (maximising expected utility) or PU (probabilistic utility) we mean.

\subsection{Probabilistic Interpretation of the Utility Function}

Let us denote with $\vec\alpha$ the set of parameters that identify a
portfolio, and with $U$ the set of parameters that
characterise our utility function model, like for instance risk aversion, investment horizon etc.
Let us assume furthermore that the
expected utility is computed with respect to a distribution characterised by
parameters, like expected excess returns for instance, that we will denote collectively with $\vec\Phi$.
The expected utility can then be written as a function
\[
	u = u(\vec{\alpha}, U, \vec\Phi).
\]
In classical asset allocation theory the prescription would be to select portfolios that
maximize the expected utility; in our framework we decided to consider the expected utility as proportional to the
logarithm of a probability measure in portfolio space, fully conditional on $U$ and $\vec\Phi$.
\begin{equation}
\vec{\alpha} \sim P(\vec{\alpha}|U,\vec\Phi) = 
      Z(\nu, U, \vec\Phi) \exp\bigg( \nu u(\vec{\alpha},U,\vec\Phi)\bigg).
\end{equation}
The symbol $\sim$ has to be interpreted as: $\vec{\alpha}$ is distribute according to, and
$Z(\nu, U, \vec\Phi)$ is a normalization constant defined by
\[
 Z^{-1}(\nu, U, \vec\Phi) = 
     \int_{D(\vec{\alpha})} [d\vec{\alpha}] \exp\bigg( \nu u(\vec{\alpha},U,\vec\Phi)\bigg) ,
\]
where $D(\vec{\alpha})$ stands for the integration domain of $\vec{\alpha}$.

The recommended portfolio $\overline{\vec{\alpha}}$, given $U$ and $\vec\Phi$, 
is defined as the expectation value of $\vec{\alpha}$:
\begin{equation}
  \overline{\vec{\alpha}}(U,\vec\Phi) =  
    Z(\nu, U, \vec\Phi)
    \int_{D(\vec{\alpha})} [d\vec{\alpha}] \vec{\alpha} 
       \exp\bigg(\nu u(\vec{\alpha},U,\vec\Phi)\bigg).
\label{popt}
\end{equation}
Since we choose to insist on a distribution to describe the portfolio, 
it is natural to identify the error associated with the
estimate of $\overline{\vec{\alpha}}$ with the standard deviation of the distribution itself.
An unbiased estimate of the standard deviation will be computed, at no extra cost, while 
computing the integral in (\ref{popt}).

The parameter $\nu$ is a constant that the theory is unable to set. Its meaning though 
is quite direct.  If we send $\nu \to 0$ we see that the density distribution 
for $\vec{\alpha}$ becomes the uniform one, all portfolios
are just as likely and the ideal one, according to the previous prescription, 
would just be evenly spread over all of the assets available. On the other 
hand, if we send $\nu \to \infty$, the ideal portfolio would just coincide with the one 
obtained by the standard MEU procedure.
For an infinite $\nu$ all of the 
measure is just concentrated about the maximum of the expected utility. 
In short, $\nu$ measures the weight that expected utility should have 
as opposed to the total noise generated by the flat  measure $[d\vec\alpha]$.
If we have a set of stationary historical data we can bootstrap from the data, build scenarios
and compute unbiased estimates of the expected utility; or, if we have
a data model and we believe that historical data are drawn from some
distribution, we can use the time series to estimate the 
distribution parameters. In both situations our confidence on the value of the expected 
utility will be in some way linked positively to the length of the available time series data set.
It seems a reasonable assumption for $\nu$ to exhibit the following 
asymptotic behaviour:
\begin{eqnarray}
	\lim_{N\to 0}\nu(N) & = & 0, \\
	\lim_{N\to\infty}\nu(N) & = & \infty,
\end{eqnarray}
where $N$ is the size of the data set.
The simplest such form is
\begin{equation}
	\nu  = \rho N^{\gamma},
	\label{nu}
\end{equation}
with $\rho $ and $\gamma$ constants strictly greater than 0.
All of the simulations carried out in this paper will have $\rho = 1$ and $\gamma = 1.$
The limit $\rho \to \infty$ will recover the standard maximisation approach. It is obviously
interesting to ask whether a more sophisticated relation between $\nu$ and $N$ could lead
to a better algorithm, in particular to one that makes a more effective use of the available
information. However, since already the simple link $\nu = N$ leads to great improvements,
these questions will be addressed in future research.  

\subsection{Bayesian Analysis and Parameter Determination}
Parameters that characterise the distribution of returns are determined 
with some degree of uncertainty that must be taken into account.
A consistent framework to do so is to accept the Bayesian point of view that 
it is not possible to infer the values of model parameters from experimental data with certainty, and to think of parameters as random 
variables themselves, 
described by a distribution. Based on the observations, we modify our view in a consistent manner with the
observed data. The result of this process will be a posterior distribution $P(\vec{\Phi}|\{R\})$,
i.e. a distribution fully conditional on the historical data $\{R\}$.

The uncertainty on the average returns must therefore play a role in the calculation
of the optimised portfolio. The Bayesian prescription to do so is to replace Eq.\ref{popt}
with the following:
\begin{equation}
 \vec{\alpha}_{PU}  = \int_{D(\vec\Phi)} [d\vec\Phi]  \overline{\vec{\alpha}}(U,\vec\Phi) P(\vec\Phi|\{R\}) \quad .
\label{bapopt}
\end{equation}

To proceed with the computation of the integral on the r.h.s of Eq.\ref{bapopt}, we need to know
the posterior distribution for $\vec{\Phi}$ that from Bayes' theorem turns out to be
\begin{equation}
	P(\vec\Phi|\{R\})  = \frac{P(\{R\}|\vec\Phi) P_0(\vec\Phi)}{P(\{R\})}.
\label{bayestheorem}
\end{equation}

The denominator $P(\{R\})$ is the unconditional distribution of the observed 
data $\{R\}$ and for our purposes but a normalisation constant, while
the two terms in the numerator represent the more interesting ones. The quantity $P(\{R\}| \vec\Phi)$
is the likelihood or probability density of the observed data subject 
to the fact that the parameters are exactly $\vec\Phi$.
The second term in the numerator of Bayes' theorem, 
$P_0(\vec\Phi )$, constitutes the a priori distribution
for the parameters, embodying thereby personal views on the expected behaviour of the 
distribution of $\vec\Phi$.  A Bayesian approach requires you to state 
explicitly what theory underlies your assumption, and the place to do
so is precisely in the choice of the prior $P_0(\vec\Phi )$. A prior should be 
chosen in accordance to our knowledge and prejudices. If we have no reason 
to believe anything at all, the prior will reflect this by assigning equal
probability to any possible configuration. It will become more and more 
decisive the stronger our convictions are rooted in background knowledge 
we have about the problem.

\section{An Explicit Posterior Distribution}
\label{post}

To proceed further we need to choose some particular data model. For the time being, and given the aims of this paper,
we resort to a classical Gaussian framework. However, it is worth noting that the selection of the data 
model could be itself a subject of Bayesian inference: we defer this extension to future research.
The posterior distribution of (\ref{bayestheorem}) can now be written out by data inspection. 
Denoting with $\vec{m}$ the average returns and with $\vec\Omega$ the covariance matrix,
the set $\{\vec{m}, \vec\Omega\}$ makes explicit what was previously referred to as $\vec\Phi$.

The likelihood term of (\ref{bayestheorem}) can be written:
\begin{eqnarray}
    P(\{R\}|\vec{m}, \vec\Omega) & = &
      \prod_{n=1}^N \frac{\exp\left(
         -\frac{(\vec{r}_n - \vec{m})^T\vec\Omega^{-1}(\vec{r}_n -\vec{m})}{2}
           \right)}
       {\sqrt{(2\pi)^J|\vec\Omega|}} \nonumber \\
 & = &  \frac{\exp\left( 
  -\frac{N}{2}\vec{m}^T\vec\Omega^{-1}\vec{m} + N\vec{m}^T\vec{\Omega^{-1}}\vec{\overline{r}} 
       - \frac{1}{2}\sum_{n=1}^N \vec{r}_n^T\vec\Omega^{-1}\vec{r}_n
           \right)}{[\sqrt{(2\pi)^J|\vec\Omega|}]^\frac{N}{2}},  \nonumber \\
\end{eqnarray}

where $J$ is the number of assets, $\vec{r}_n$ the n-th observations vector, and: 
\[
 \overline{\vec{r}} = \frac{1}{N}\sum_{n=1}^N \vec{r}_n.  
\]
We will choose a prior for $\vec{m}, \vec\Omega$ of the form:
\begin{equation}
    P_0(\vec{m}, \vec\Omega) = P_0(\vec{m}| \vec\Omega) P_0(\vec\Omega),
\end{equation}
where the average conditional distribution is chosen as a normal,
\begin{equation}
    \label{beta}
    P_0(\vec{m}| \vec\Omega) \simeq  \exp\left(
       -\frac{\beta}{2}(\vec{m}-\vec{\chi})^T\vec\Omega^{-1} (\vec{m}-\vec{\chi})
        \right).
\end{equation}
The vector $\vec\chi$ is the view we hold, consistent with our background knowledge, of the central point
of the distribution of average returns, while $\beta$ is a hyper-parameter 
that the theory cannot fix; it controls the width of the distribution and  we 
will soon see a possible interpretation.

The prior for the covariance matrix is the inverse Wishart:
\begin{equation}
    P_0(\vec\Omega) \simeq |\vec\Omega|^{-\frac{h+J+1}{2}}\exp\bigg[\vec\Omega^{-1}\vec{\Sigma}_{\Omega}\bigg],
    \label{h}
\end{equation}
where $h$ is once again another hyper-parameter and $\vec\Sigma_{\Omega}$ is our view.

Putting all together, we have:
\begin{equation}
    P_0(\vec{m}, \vec\Omega|\{R\}) = P_0(\vec{m}|\vec\Omega, \{R\}) P_0(\vec\Omega|\{R\}),
\end{equation}
where:
\begin{eqnarray}
    P_0(\vec{m}|\vec\Omega, \{R\}) & \simeq  &
    \exp\bigg(-\frac{N(1+\kappa)}{2}(\vec{m}-\vec{M})^{\ T}\vec\Omega^{-1}(\vec{m} - \vec{M})\bigg)\\
    P_0(\vec\Omega|\{R\}) & \simeq & |\vec\Omega|^{-\frac{h+N+J-1}{2}} \exp\bigg(
     - \frac{N+h}{2}Tr\left[\vec\Omega^{-1} \vec{A}\right]
           \bigg) \\
    \vec{M} & = & \frac{\vec{\overline{r}} + \kappa\vec\chi}{1+\kappa}, \\
    \kappa & = & \frac{\beta}{N}, \\
    \vec{A}   & = &  \frac{h\vec{\Sigma}_{\Omega}}{N+h} +  \frac{N\vec\Sigma}{N+h} 
      + \frac{\kappa(\vec{\overline{r}}-\vec\chi)(\vec{\overline{r}}-\vec\chi)^{\ T}}{(1+\kappa)(N+h)} \\
    \vec\Sigma &=& \frac{1}{N}\sum_{n=1}^N (\vec{r}_n - \vec{\overline{r}})(\vec{r}_n-\vec{\overline{r}})^{\ T}.
\end{eqnarray}

All the details of the computation can be found in \ref{priors}.

It seems natural to view $\kappa$ and $h$ as a simple way to measure the degree of 
confidence we have in our views as opposed to
the indications stemming from historical data. If we hold a view but we think that 
observed data should weigh more in our decision process, then we would choose small values 
for $\kappa$ and $h$. Note that in the limit $\kappa \to 0$, we would recover,
for the average return, a totally non-informative 
prior that assigns equal probability to any possible value of $\vec{m}$. 
A strong view is represented by a large $\kappa$ and large $h$.  
In the limit $\kappa\to\infty$  and $ h \to \infty$, the posterior distribution would be centred 
about our views regardless of the historical data, and the width of the distribution would tend to $0$. 

\section{Numerical Integration}
 \label{num}
\subsection{Markov Chain Monte Carlo Integration}
The integral in Eq.(\ref{bapopt}) is easily carried out by Markov Chain Monte Carlo (MCMC) 
integration \cite{Gilks96, Polson02}. Since the
probability distribution for $(\vec{m}, \vec\Omega)$ is independent from the distribution of
$\vec{\alpha}$, an algorithm that generates the Markov Chain capable of
yielding the correct distribution is as follows:
\begin{description}
\item[{\bf Step 1}] Sample $\vec\Omega$ from the inverse Wishart probability density function (p.d.f.):
\begin{equation}
     |\vec\Omega|^{-\frac{h+N+J+1}{2}}\exp\bigg[(N+h)Tr\big[\vec\Omega^{-1} \vec{A}\big]\bigg].
\end{equation}
This can be achieved by generating $N+h$  J-dimensional arrays $\vec{x}_i, i=1, \dots, N+h$
distributed according to
\begin{equation}
    \vec{x} \sim N(0, \vec\Omega^{-1}),
\end{equation}
and setting
\begin{equation}
    \vec{A} = \frac{1}{N+h}\sum_{i=1}^{N+h}\vec{x}_i\vec{x}_i^{\ T}.
\end{equation}
\item[{\bf Step 2}] Holding fixed the sampled $\vec\Omega$, sample $\vec{m}$ from the p.d.f.:
\begin{equation}
     \exp\left( -\frac{N(1+\kappa)}{2}(\vec{m}-\vec{M})^T\vec\Omega^{-1}(\vec{m}-\vec{M}) \right).
\end{equation}
\item[{\bf Step 3}] Holding fixed the values for $(\vec{m}, \vec\Omega)$, sample $\vec{\alpha}$
from the p.d.f. 
\begin{equation}
    \exp\bigg( N E\big[u(\vec{\alpha}, U, \vec{m}, \vec\Omega)\big]\bigg).
\end{equation}
\end{description}
Details of the algorithm and the proof that produces an unbiased 
estimate of the integral in the r.h.s of Eq.(\ref{bapopt})
can be found in \ref{integration}.

Steps 1 and 2 do no present any problem since we know how to sample from those p.d.f.s. Step 3 is somewhat more complex.
We do not know how to sample directly from that p.d.f., and we are forced to devise a Markov chain that relaxes to
the desired distribution. After several experiments with variations of the Metropolis-Hastings, we resorted to
an implementation of the ''Hybrid Monte Carlo" method. Once relaxation has been achieved we can run the Markov chain 
for few more steps in order to perform measurements. Relaxation or thermalisation is not a trivial issue but a thorough
discussion of the problems involved would bring us too far from the subject of this paper. We choose to defer this 
discussion to a forthcoming paper focussing on the implementation of the numerical integration scheme.

\subsection{The Utility Function}
The selection of a good utility function is not the subject of this paper, nor is it particularly
relevant for our results. 
Whenever the return distribution is assumed to be normal, as in our framework, the explicit 
solutions of all the utility funtions are but a combination of first and second distribution moments.
Still, a non-trivial difference arises when standard deviation terms are included,
since they are able to generate a time horizon effect, i.e. an effect that favours less risky assets on the short 
range and turns on risky ones, with higher returns, on the long range. 
We are aware of the academic debate on this topic,
testified by a considerable amount of related literature \cite {CampbellViceira01,CampbellViceira02,KritzmanRich,Samuelson94,SThorley}, and
we believe this to be a desirable feature for a
utility function. The probability for the 
riskier assets to outperform the less risky ones, in fact, approaches one asymptotically with time,
being it the error function of the ratio between mean and standard deviation, which 
grows with the square root of time. 

However, utility functions of standard use in financial economics (such as those who 
exhibit Constant Relative Risk Aversion) do not fall in this category. 
The standard deviation terms are directly
related to non-regular utility functions that measure risk
with the concepts of, say, Value at Risk, Loss Probability etc., i.e. with the so-called downside risk measures \cite{Bouchaud}.
In this way risk is measured by the expected amount by which a specified target is not met: this 
might better describe
how the investor perceives risk, as documented by
results from behavioural finance \cite{Hirsh}, and is more in line with some recent ALM practice.

For these reasons we employed the following expected utility function, drawn from the article of
Consiglio \etal \cite{Consiglio-et-all01}:

\begin{equation}
E\bigg[u(\vec{\alpha}, L, \lambda, T )\bigg]  = 
  \sum_{n=1}^{N_T} \Delta t 
   [E(U(n\Delta t))-\lambda E(D(n\Delta t))]   ,
\end{equation}

where $U(n\Delta t)$ and $D(n\Delta t)$ are the upside and downside, respectively, of the portfolio return at time $n\Delta t$ 
against a fixed target return $L$, and $\lambda$ is a weight indicating the investor risk aversion. 
The time horizon T is built out of $N_T$ intermediate 
time intervals $\Delta t$ such that $T = N_T\Delta t$, is a sequence of $N_T$ values for
$\omega(n\Delta t), \quad n=1,\ldots,N_T$. 
The model takes a "target-all time" view, and the allocation is such that staying as close to the target return trajectory at all times is the primary concern.
A risk averse investor will want to keep as far as possible from target return under-performance 
situations, and will favour paths close to the target line.

Modelling the distributions for the single period log-return $\vec\omega$  
with the normal $N(\vec{m}, \vec\Omega)$,

\begin{equation}
              \vec\omega \sim
      \exp\left(
         -\frac{(\vec\omega- \vec{m})^T\vec\Omega^{-1}(\vec\omega-\vec{m})}{2}
           \right)
       \frac{1}{\sqrt{(2\pi)^J|\vec\Omega|}},
\label{returndist}
\end{equation}

we obtain by straightforward (tedious) Gaussian integration an explicit expression for the utility function:

\begin{equation} 
    E\bigg[u(\vec{\alpha},L,\lambda,T)\bigg]  
          = \sum_{n=1}^{N_T}\Delta t\bigg[ 
          n\Delta tM f_2(n\Delta t) - \sqrt{n\Delta t}S f_1(n\Delta t) 
         \bigg],
\label{expectedutility}
\end{equation}
where:
\begin{eqnarray}
     f_1(t) & = & (\lambda - 1)\frac{e^{-t\eta^2}}{\sqrt{2\pi}}  \\ 
         f_2(t) & = & \frac{1+\lambda}{2} - \frac{\lambda-1}{2}\mbox{erfc}(\sqrt{t}\eta)  \\
    \eta  & = &  \frac{M}{2S} \\
        M  & = & \vec{\alpha}^T\cdot\vec{m} - L \\
        S^2 & = & \vec{\alpha}^T\cdot\vec\Omega\cdot\vec{\alpha}.
\end{eqnarray}

As expected, the explicit solution for this specific form of utility is a function of the portfolio 
mean, variance and -most importantly- standard deviation. It
incorporates a competing effect 
between average return and standard deviation with different
time scaling properties: the standard deviation's contribution is proportional to $\lambda - 1$,
and can be traced back to the imperfect cancellation between positive and 
negative deviations from ideal line. We thus obtain the desired dependency of the optimal portfolio on the chosen
time horizon: the longer the horizon (ceteris paribus), the more aggressive the optimal allocation.
\section{Empirical Results}
\label {results}

In this section the performance of our proposed PU method will be analysed in various contexts: first 
its consistency and speed of convergence will be tested and compared to the MEU optimisation with the 
help of artificial data generated by a known multivariate normal distribution. Afterwards the performance
of both prescriptions will be reviewed by means of historical time series data. As mentioned before, 
the sensitiveness towards the risk aversion parameter $\lambda$ of both selection procedures will also be 
evaluated. Finally, the effect of incorporated personal views is illustrated, and the degree of confidence
associated with an ``optimised'' portfolio is discussed.

Historical data used to infer distribution parameters consists of 8 monthly indexes covering the period
from January 1988 to January 2002. In Tabl.\ref{listoftitles} we 
show the list of titles employed; this set of data will be referred to as full sample in the following.

\begin{table}[ht]
\caption{ \label{listoftitles} List of assets employed.
The full set of tha data goes from Jan 1988, to Jan 2002.
The used acronyms have the following meaning: 
 {MSCI = Morgan Stanley Capital Index,}
 {JPM = JPMorgan Index,}
 {ML = Merrill Lynch Index .}
 Data source: Datastream.
 Data types	: Price Index for equities, Total Return Index for bonds.
 All the samples are in local currency, unadjusted for inflation.
 The index titles refer to the Datastream mnemonics.
}
\vskip .2cm
\begin{indented}
\item[]\begin{tabular}{@{}ll}
\br 
Assets & Description \\
\mr 
MSNAMR  & MSCI North America Equity    \\ 
MSPACF  & MSCI Pacific Equity          \\ 
MSEROP  & MSCI Europe Equity           \\ 
JPMUSU  & JPM  US Government Bond      \\ 
JPMJPU  & JPM  Japan Government Bond   \\ 
JPMEIL  & JPM  Europe Government Bond  \\ 
MLHMAU  & ML   US Corporate High Yield \\ 
JPEC3M  & JPM  Euro Cash               \\ 
\br 
\end{tabular}
\end{indented}
\end{table}


\subsection{Simulation with Artificial Data}

\begin{figure}[ht]
\begin{center}\scalebox{.6}{ \epsfbox{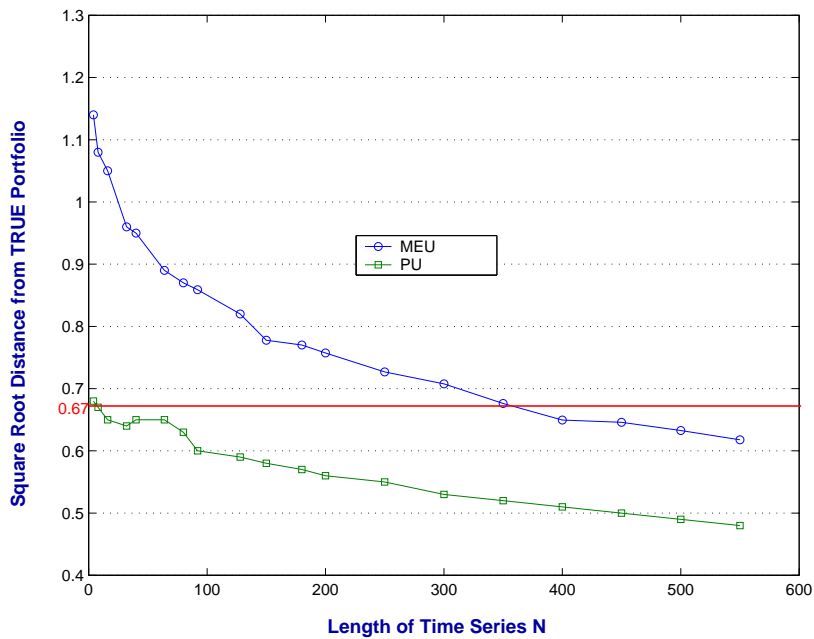}}
\caption{\label{N_figure} Distance from TRUE Portfolio w.r.t. sample size: for each sample lenght,
average euclidean distance
of the MEU and PU portfolios (resulting from 1000 independent samples) from the
known true optimal allocation. The benchmark value for a randomly chosen portfolio is represented
by the straight line.} \end{center}
\end{figure}

We first investigate the performance of our proposed PU method by using artificial data to simulate finite sample
behaviour. For the testing we assume the true return distribution to be a multivariate Gaussian, characterised by the 
parameters estimated from the full historical sample. 
From this distribution we generate 1000 independent samples of various fixed lengths.
For any given sample length and a fixed parameter set (L=$5\%$ per yr, T=$1$ yr, and $\lambda=3$), 
we then calculate the average Euclidean
distance of both the MEU and PU 1000 optimal portfolios from the ``truly''
optimal allocation, that we can determine exactly from the parameters of the assumed ``true" distribution.    
In Fig.\ref{N_figure} we have plotted the results of this exercise, together with a straight line showing the average 
distance of a randomly chosen portfolio from the ``true'' one\footnote[1]{for a derivation of this value refer to \ref{Random Portfolios}}.
Our PU method clearly outperforms the MEU procedure, for it is always closer to the true allocation and below
the random-choice threshold. 
The picture well illustrates the extreme sensitiveness of the MEU procedure 
to the input data; for a great distance from a benchmark portfolio when averaging over 1000 samples can only 
be explained by a great variability in the portfolio composition over the different samples.

Asymptotically, for $N\to\infty$, the return distribution parameters are determined with quasi-certainty, 
and we consequently  recover the ``true' optimal portfolio, thereby verifying the consistency
of both approaches. In Tabl.\ref{neffect} we present evidence for this, reporting the allocations for $N=32000$ 
observations.

However, for the classical MEU optimisation the speed of convergence looks worryingly slow when considering 
typical lengths of time series data used in asset allocations by practitioners.
Indeed, for the chosen set of parameters one observes that up to a threshold of more than 350 monthly observations, corresponding to a 
data sample of almost 30 years, 
the knowledge gained from data is actually insufficient for selecting any but the equally-weighted portfolio!
This nicely illustrates the real risk of estimation errors completely overwhelming the value of information 
contained in the data. A restriction to very long data sets could seem a solution (provided data is available),
but then one could object
again by referring to the well known non-stationarity exhibited by financial time series.
On the other hand, our PU prescription manages to stay always below the random portfolio threshold line, 
although coming very close to it when observations are scarce, thereby justly reflecting a situation in 
which data is not sufficient to justify very ``particular'' portfolios. 

 \begin{table}[ht]

 \vskip .2cm
 \begin{indented}
 \item[]
 \begin{tabular}{@{}lcccccc}
 \br
  {\bf ASSET } & {\bf  PU  } & {\bf MEU } \\
  \mr
  MSNAMR   & $0.04 \pm  0.01$ & $0.04$  \\
  MSPACF   & $0.00 \pm  0.00$ & $0.00$  \\
  MSEROP   & $0.00 \pm  0.00$ & $0.00$  \\
  JPMUSU   & $0.02 \pm  0.02$ & $0.02$  \\
  JPMJPJ   & $0.00 \pm  0.00$ & $0.00$  \\
  JPMEIL   & $0.28 \pm  0.08$ & $0.27$  \\
  MLHMAU   & $0.04 \pm  0.02$ & $0.04$  \\
  JPEC3M   & $0.62 \pm  0.10$ & $0.63$  \\ 
 \br
 \end{tabular}
 \end{indented}
  \caption{\label{neffect}
 Optimal portfolio for N=32000. The investment parameters are constant and set to
  $\lambda=3$, T=1 yr, L=$5\%$. 
 }
 \end{table}

\subsection{Backtesting on Historical Data}

Based on the full sample historical data and for some chosen set of parameters, we have performed some back testing, in the form
of ``what if'' investment scenarios. To this end, we used 5 years rolling windows for estimation and 3 years rolling windows for
out-of-sample testing, together with two larger samples. We measure each portfolios' hypothetically achieved performance\footnote[2]{Of course, such an ex
post performance verification for some (by us) chosen set of time series and investor profile does not allow
to draw definite conclusions; it is merely meant to support and illustrate the more important results
from the above section.}.  
In Tabl.\ref {back} we present back testing results for different samples; for each sample, we have selected an ``average'' risk attitude, 
and computed a unique performance indicator (Sharpe Ratio), 
neglecting the behaviour at intermediate intervals. Results vary over the examined
samples: until the first half of the '90s, when financial series exhibited more stable patterns, the MEU procedure achieves 
better performances, while for more recent samples it is the PU that outperforms the MEU optimisation. 
Using longer samples advantages the PU procedure.

 \begin{table}[h]
 
 \vskip .2cm
 \begin{indented}
 \item[]
 \begin{tabular}{@{}cccc}
 \br
  {\bf Estimation Sample } & {\bf  Out-of-Sample  } & \multicolumn{2}{c} {\bf Sharpe Ratio } \\
  \cline{3-4} & & MEU & PU \\
  \mr
  $1988-1992$   & $1993-1996$ & $3.17$ & $2.42$  \\
  $1989-1993$   & $1994-1997$ & $2.64$ & $2.44$  \\
  $1990-1994$   & $1995-1998$ & $2.63$ & $2.37$  \\
  $1991-1995$   & $1996-1999$ & $2.10$ & $2.19$ \\
  $1992-1996$   & $1997-2000$ & $1.61$ & $1.77$ \\
  $1993-1997$   & $1998-2001$ & $0.84$ & $0.89$ \\
  $1994-1998$   & $1999-2002$ & $-0.23$ & $-0.15$ \\
  $1988-1995$   & $1996-2002$ &	$0.71$ & $0.78$ \\
  $1988-1997$   & $1998-2002$ & $0.20$ & $0.31$ \\ 
 \br
 \end{tabular}
 \end{indented}
 \caption{\label{back}
  Back-testing. 
 }
 \end{table}

\subsection{Sensitiveness Towards Risk Aversion Parameter $\lambda$}
As a second empirical investigation, we examine the algorithms' stability for a given portfolio profile 
(L=$5\%$ per yr, T= $1$ yr). As previously stressed, all 
expected-utility based procedures suffer from the presence of a risk aversion parameter, dimensionless and un-settable from
theory. As a measure of instability, it seems then natural to compare the sensitivity to $\lambda$ for both the MEU optimisation 
prescription and the PU method we propose in this paper. 
Specifically, we examine the behaviour of a diversification indicator, i.e. an indicator that measures the degree of concentration
within a portfolio, and consequently
allows to identify the range of the parameter that mostly affects the portfolio composition. 

The simplest of such a - as Bouchaud \etal\cite{Bouchaud2} put it - entropy-like measure
is the quantity: 
\begin{equation}
Y = \sum_{j=1}^J  \alpha_j^2  \quad.
\end{equation}
which ranges from $\frac{1}{J}$ ($J$=number of assets= 8 here), when the portfolio is totally diversified (evenly spread), to $1$ in case of 
complete concentration on one asset.

\begin{figure}[ht]
\begin{center}\scalebox{.6}{ \epsfbox{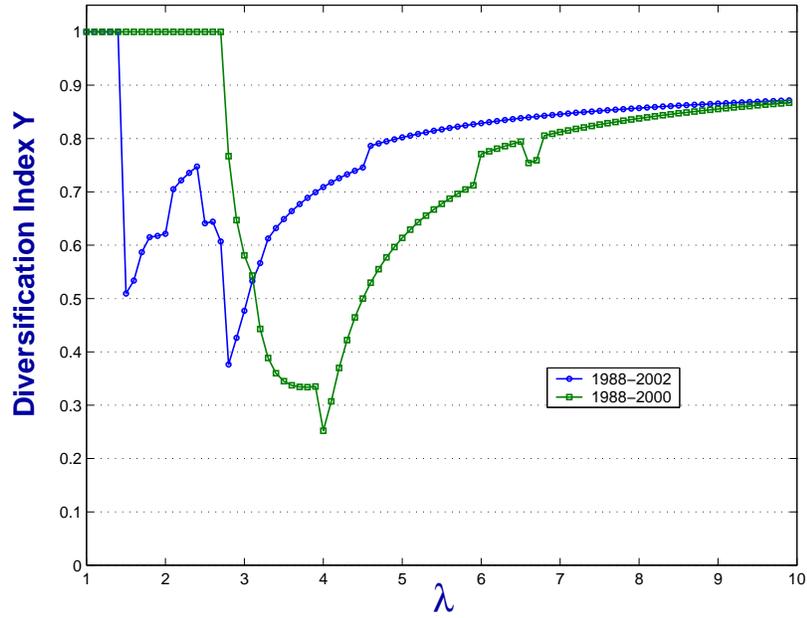}}
\caption{\label{lambdaOpt} Portfolio diversification w.r.t. risk aversion $\lambda$ and data sample: MEU Procedure} \end{center}
\end{figure}

\begin{figure}[ht]
\begin{center}\scalebox{.6}{ \epsfbox{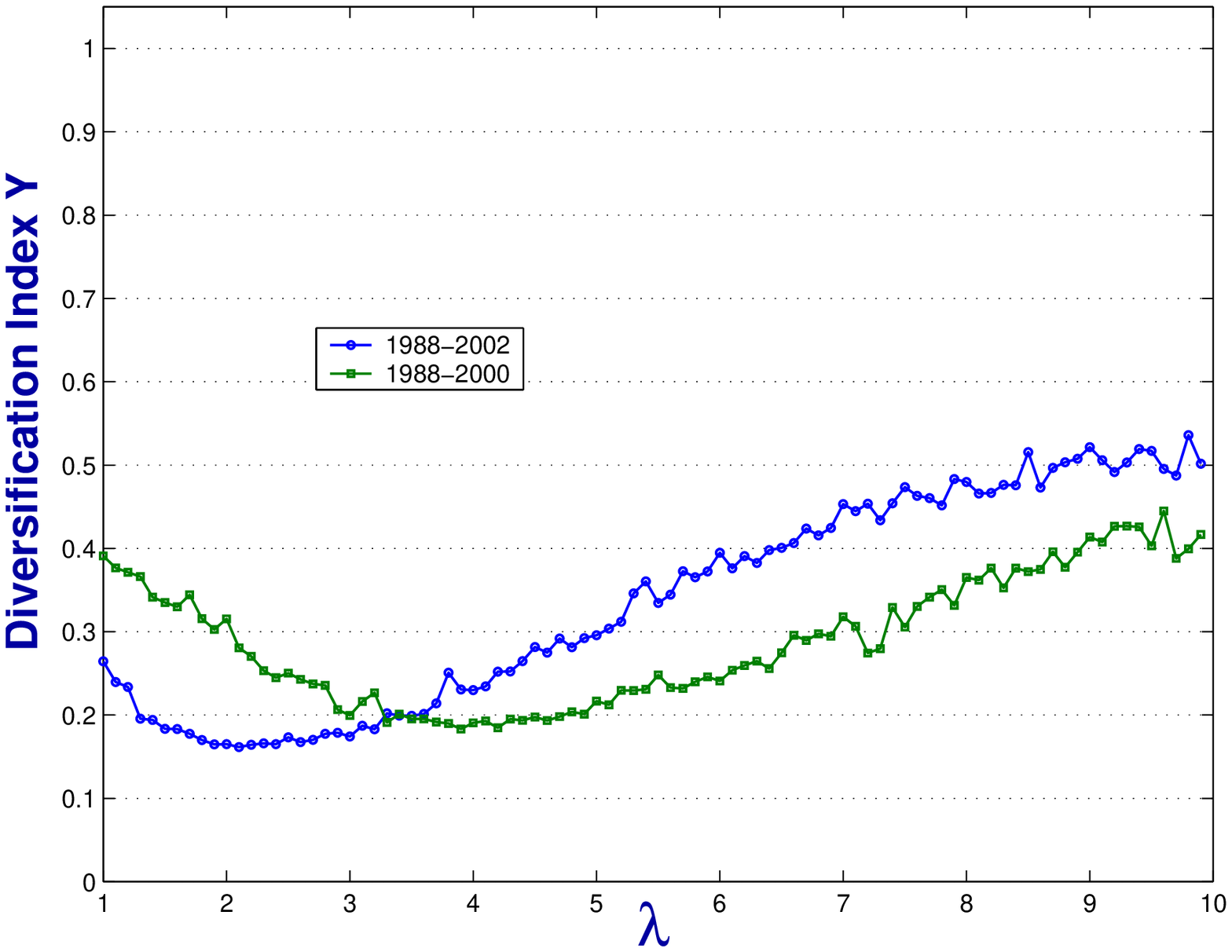}}
\caption{\label{lambdaMCMC} Portfolio diversification w.r.t. risk aversion $\lambda$ and data sample: PU Procedure} \end{center}
\end{figure}

In Fig.\ref{lambdaOpt} and Fig.\ref{lambdaMCMC} we report the behaviour of Y with respect to $\lambda$ for two different data 
samples, the full sample and a slightly restricted 1988-2000 one.
Looking at the MEU graph in Fig.\ref{lambdaOpt}, the behaviour appears very erratic and the significant range of $\lambda$ 
restricted to a relatively small interval, meaning that small changes in $\lambda$ can produce
large modifications in the portfolio composition. 
Indeed, if we look at Tabl.\ref{tabellalambda}, we can observe 
how the portfolio composition changes as $\lambda$ moves from $2.6$ to $3.0$, to the point that the portfolios 
are totally twisted around. 
This is certainly not reassuring, given that $\lambda$  is only
loosely tied to investor's risk aversion, and its setting is not without uncertainties.
Back to Fig.\ref{lambdaOpt}, what strikes even  more is what happens when we look at the results for a different data sample, in this case shortened 
by the last two years of observations: 
the curve decidedly shifts to the right, and consequently the relevant range of $\lambda$ does the same, leading to dangerous
risk profile mis-identifications, and forcing to re-calibrate (with all the associated uncertainties) the values of $\lambda$ 
basically each time new historical observations are added to the sample.
                                     
Coming now to the PU model, for which results are shown in Fig.\ref{lambdaMCMC}, 
the diversification indicator displays a very different pattern: it indicates a 
more conservative overall
behaviour, with values closer to the lower bound of $\frac{1}{8}$.
It never concentrates all the weights on a single asset, not even for the risk-neutral ($\lambda=1$) case.
In Tabl.\ref{tabellalambda} we can see data from our Bayesian PU approach: the variations in the portfolios 
induced by the different $\lambda$'s are now hardly noticeable.
Most importantly, $Y$ exhibits a smooth 
pattern. This reduces the danger of mis-settings of $\lambda$ and its sensitiveness on the chosen sample; 
for different data sample, in fact, the curve shifts but remains rather similar, 
leaving unaffected the significant range of $\lambda$ . 

 \begin{table}[ht]

 \vskip .2cm
 \begin{indented}
 \item[]
 \begin{tabular}{@{}lcccccc}
 \br
  {\bf ASSET } & \multicolumn{3}{c}{\bf  MEU } & \multicolumn{3}{c}{\bf PU } \\
 \cline{2-7}&$\Lambda_1$&$\Lambda_2$&$\Lambda_3$&$\Lambda_1$&$\Lambda_2$&$\Lambda_3$ \\
  \mr
  MSNAMR   & $0.11$ & $0.06$ & $0.04$ & $0.16 \pm  0.25$ &$0.14 \pm  0.23$ & $0.13 \pm 0.22$ \\
  MSPACF   & $0.00$ & $0.00$ & $0.00$ & $0.01 \pm  0.01$ &$0.01 \pm  0.01$ & $0.01 \pm 0.01$ \\
  MSEROP   & $0.01$ & $0.00$ & $0.00$ & $0.08 \pm  0.15$ &$0.07 \pm  0.14$ & $0.06 \pm 0.12$ \\
  JPMUSU   & $0.00$ & $0.02$ & $0.00$ & $0.15 \pm  0.19$ &$0.15 \pm  0.18$ & $0.14 \pm 0.18$ \\
  JPMJPJ   & $0.00$ & $0.00$ & $0.00$ & $0.03 \pm  0.04$ &$0.03 \pm  0.04$ & $0.03 \pm 0.04$ \\
  JPMEIL   & $0.79$ & $0.43$ & $0.28$ & $0.23 \pm  0.24$ &$0.23 \pm  0.23$ & $0.23 \pm 0.23$ \\
  MLHMAU   & $0.10$ & $0.06$ & $0.04$ & $0.15 \pm  0.20$ &$0.14 \pm  0.19$ & $0.13 \pm 0.18$ \\
  JPEC3M   & $0.00$ & $0.43$ & $0.64$ & $0.20 \pm  0.26$ &$0.23 \pm  0.28$ & $0.26 \pm 0.30$ \\ 
 \br
 \end{tabular}
 \end{indented}
   \caption{ \label{tabellalambda}
   Optimal portfolio with respect to $\lambda$. For all columns the time horizon T is one year
   and expected return $5\%$ per year. The whole sample (1988-2002) is considered.
   The parameter $\lambda$ instead is set to:
   $\Lambda_1:\lambda=2.6$, 
   $\Lambda_2:\lambda=2.8$ and 
   $\Lambda_3:\lambda=3$. 
 }
 \end{table}

\subsection{Confidence Associated with Identified Portfolio}

While the MEU optimisation procedure places infinite faith on the distribution as 
determined through simplistic inspection of historical data, and makes no allowance for imperfect knowledge, 
the Bayesian PU approach has this naturally built in. The decidedly more conservative allocations that 
are manifested in our more balanced portfolios reflect the fact that we 
do not exactly know what the true distribution is, and thus we try to protect ourselves against 
situations in which actual distributions are rather different from the ones we 
are invited to deduce from historical data. Another hint for the relative smallness of the full sample
comes from the observation of the portfolio's standard errors quoted in Tabl.\ref{tabellalambda}. 
They are of the same magnitude 
of the average value (=weight) of the asset, indicating that we should not take too seriously a prediction of 
$14\%$ as opposed to a $20\%$. From this table we can safely conclude which assets should not be in our 
portfolio, while when it comes to the single best way to distribute the others we can, at best, be only suggestive.
The confidence intervals might appear too large, but they are just 
another confirmation ex post of the inner consistency of our formalism: from Fig.\ref{N_figure} we know that the
square distance of our estimated portfolio from the "true" portfolio should be
of the order of $d^2$=$0.57^2$=0.32 in this specific case (full sample, 180 observations). 
As a rough test, we might
double check this number by summing the squares of the PU calculated
standard errors as displayed in Tabl.\ref{tabellalambda}, where we find 0.27, which results very
well compatible within this rough consistency check.

Since MEU optimisation completely trusts its return distribution parameters as
estimated by available data, it naturally misses a means to characterise the
degree of confidence to be attached to its "optimal" portfolio. However, from
our finite sample tests as shown in Fig.\ref{N_figure} we are able to give a rough estimate
of the mean standard error of every asset's weight: 0.77 is the average distance
from the true portfolio in the case of $N=180$ (our full sample), and therefore $0.77$/$\sqrt{8}$= $0.27$
should be a reasonable estimate of the error, ignored by the MEU formalism, but
definitely a reality which should not be denied.

\subsection{Effects of Incorporated Personal Views}

 In Tabl.\ref{tabellaviews} it is illustrated what happens when we express personal views for the distribution moments.
 In the first column we report the portfolio allocation for neutral views. In the second one, we have incorporated views 
 only on the mean values of the 
 equity indexes: we postulate a very optimistic scenario, with an annual average return of $11\%$ for MSNAMR and MSEROP, 
 and of $15\%$ for MSPACF.
 We attach a rather strong degree of confidence to this personal view, setting $k=10$ (i.e.$\beta=10N$). 
 The results are in line with the previously expressed views: MSPACF, that
 was not selected in the neutral-views scenario because of the poor performance over our historical data sample, is now 
 the most over-weighted asset class, since it was modified by our strong expectation. The portfolio errors drop consequently, because 
 of the confidence degree attached to the views.
 In column three we repeat the exercise with views on variances. We express views, again only on equities, based on the implied volatilities
 inferred from proxy indexes options with two years expiration. The values are quite large if compared to the historical ones 
 (annualised implied volatilities: 
 MSNAMR $25\%$ MSPACF $26\%$, and MSEROP $29\%$),
 reflecting the market sentiment for the near future. We set $h=N$, so that the resulting 
 variances are the average between the historical and the implied ones. As expected, the resulting asset allocation is
 more conservative, and, given the errors size, almost all asset classes are included in the portfolio.

  \begin{table}[ht]

 \vskip .2cm
 \begin{indented}
 \item[]
 \begin{tabular}{@{}lccc}
 \br
  {\bf ASSET } & \multicolumn{3}{c}{\bf  PU  }  \\
 \cline{2-4}&$k=0,h=0$&$k=10,h=0$&$k=0,h=N$ \\
  \mr
  MSNAMR   & $0.39 \pm  0.41$ &$0.23 \pm  0.24$ & $0.20 \pm 0.31$ \\
  MSPACF   & $0.00 \pm  0.03$ &$0.58 \pm  0.30$ & $0.06 \pm 0.16$ \\
  MSEROP   & $0.19 \pm  0.34$ &$0.07 \pm  0.13$ & $0.15 \pm 0.26$ \\
  JPMUSU   & $0.08 \pm  0.22$ &$0.01 \pm  0.05$ & $0.12 \pm 0.24$ \\
  JPMJPJ   & $0.00 \pm  0.01$ &$0.00 \pm  0.00$ & $0.07 \pm 0.18$ \\
  JPMEIL   & $0.18 \pm  0.32$ &$0.08 \pm  0.17$ & $0.15 \pm 0.26$ \\
  MLHMAU   & $0.15 \pm  0.31$ &$0.03 \pm  0.10$ & $0.16 \pm 0.29$ \\
  JPEC3M   & $0.01 \pm  0.06$ &$0.00 \pm  0.00$ & $0.10 \pm 0.25$ \\ 
 \br
 \end{tabular}
 \end{indented}
  \caption{ \label{tabellaviews}
 Incorporating views: portfolio selection for neutral views (column 1), views on means (column 2) and views
 on variances (columns 3). The parameters set are constant and equal to
  $\lambda=3, T=5 years, L=11\%$. 
 }
 \end{table}

\section{Conclusion and final remarks}
\label{concl}
The purpose of this work was to address and improve some of the well known weaknesses of
portfolio selection by maximising expected utility.
We have pointed out that seeking  and settling on an extremum of a utility function is equivalent to claim
absolute knowledge of the parameters governing the distribution of average returns.
While theoretically this is never the case, historical data might at best offer partial support
to our selection process, for which we attempted to provide a unified framework.
The approach presented here takes into account parameter uncertainty 
and greatly reduces the instability of results common in standard optimisation procedures.
We achieved this by employing a different interpretation of the utility function, and by endorsing a Bayesian framework approach.
In doing so one benefits from several advantages: the framework provides a consistent way to
account for uncertainty and, whenever we hold views, we can readily introduce them. Moreover, the standard
error calculated easily for the recommended portfolio gives a good idea about the degree of certainty offered
by the available historical data. We have tested the proposed method against 
traditional expected utility maximisation, using artificial data to simulate finite-sample behaviour, and have shown superior 
performance of our method as compared  to the simplistic optimisation. This picture was reinforced 
when backtesting with historical data. 
We also managed to significantly improve the intrinsic instability with respect to the risk-aversion parameter
(lack of continuity) that plagues all maximisation approaches. 

As for future lines of research, we might be interested in relaxing the normality assumption, for instance by modelling the data with a mixture of Gaussian distributions:
in Section \ref{post} we hinted that the selection of the data 
model could itself be a subject of Bayesian inference.
Additionally, there were some occasions in which our theory led to parameters or hyper-parameters that could
easily be determined in their asymptotic behaviour, but whose value on the intermediate range was not clear (
$\nu$ in Eq.\ref{nu}, $h$ in Eq.\ref{h}, $\beta$ in Eq.\ref{beta}). Especially $\nu$,
the smoothing parameter within our probabilistic utility certainly has an important influence on the overall
performance of our method; it would therefore be interesting to ask  whether a more sophisticated prescription
than the by us employed $\nu = N$ could lead to an enhanced overall performance, i.e., in particular, to 
a faster convergence towards any ``true'' optimal portfolio.

\ack{
The authors thankfully acknowledge the exchange of opinions 
with T. Kennedy and A. Pellisetto for useful insight on the MCMC integration.}

\newpage
\appendix

\section{Random Portfolios}
\label{Random Portfolios}
Let us denote with $\overline{\vec{p}}$ a particular portfolio. It is instructive to ask
what would be the average square distance from it if we were to draw random portfolios.
In this context 'random' means portfolios drawn uniformly from the hyperplane
characterised by the equality constraint:
\[ \sum_{j=1}^J p_j = 1 \]
and the $J$ inequality constraints 
\begin{eqnarray}
 p_i & \ge & 0, \\
   & \ldots &  , \\
 p_j & \ge & 0.
\end{eqnarray}

The expected value of a function with respect to this measure is defined as:
\begin{equation}
\fl
 E[O] \equiv Z\int_0^1 dp_1 \int_0^{1-p_1}dp_2 \ldots \int_0^{1-\sum_{j=1}^{J-2}p_j}dp_{J-1}
  O(p_1,\ldots,p_{J-1}, 1-\sum_{j=1}^{J-1}p_j),
\end{equation}
where $Z$ is a normalisation constant defined by
\[
  1 = Z\int_0^1 dp_1 \int_0^{1-p_1}dp_2 \ldots \int_0^{1-\sum_{j=1}^{J-2}p_j}dp_{J-1}.
\]
Recalling the result 
\begin{eqnarray}
\fl
\int_0^1dp_1 \int_0^{1-p_1}dp_2 \ldots \int_0^{1-\sum_{j=1}^{J-2}p_j}dp_{J-1}
  p_1^{a_1}\ldots p_{J-1}^{a_J} = 
  \frac{\prod_{j=1}^J \Gamma(1+a_j)}{\Gamma(\sum_{j=1}^J(1+a_j))},
\end{eqnarray}
we have:
\begin{eqnarray}
	Z & = & \Gamma(J) \\
	E[p_i] & = & \frac{1}{J} \\
	E[p_ip_j] & = & \frac{1}{J(J+1)}\quad\mbox{for}\quad i \ne j \\
	E[p_i^2] & = & \frac{2}{J(J+1)} 
\end{eqnarray}
and the average square distance is given by:
\begin{eqnarray}
 E[\sum_{j=1}^J (p_j - \overline{p}_j)^2] = \sum_{j=1}^J\overline{p}_j^2 - \frac{2}{J(J+1)}.
\end{eqnarray}

\section{Notation and detailed balance}

In this appendix we introduce the basic concepts of Markov Chain Monte Carlo for
the sole purpose of reviewing notation and fundamental results. The field is too vast 
to get into any depth within a few pages. The interested reader might refer to the
literature on the subject, like for instance Gilks, Richardson and Spiegelhalter \cite{Gilks96},
and references cited therein.

\subsection{Monte Carlo Integration}
Let $P(x)$ be a probability density for a random variable x.  If we draw samples $\{x_i, i = 1, \ldots, n\}$
from $P(x)$, we can evaluate the average $E\left[g(x)\right]$ of an arbitrary function $g(x)$
\begin{equation}
E\left[g(x)\right]  \approx \frac{1}{n}\sum_{i=1}^n g(x_i).
\label{montecarlointegration}
\end{equation}

\subsection{Markov Chains}
Let $ T(x|y)$ a matrix describing the probability to get 'x' if we have 'y', then we can generate a 
Markov Chain (sequence of random variables) $x_1, x_2, \ldots, x_t, \ldots$ such that the probability to get $x_t$
is described by $T(x_t|x_{t-1})$.

If $T(x|y)$ satisfies the equation:
\begin{equation}
	P(x) = \int dy T(x|y) P(y),
\label{kerneleq}
\end{equation}
the following theorem holds:

\par{\bf Theorem 1: }{\it
Let $\{x_0, x_1, \ldots, x_t, \ldots\}$ be the Markov Chain generated with transition probability
$T(x|y)$. If $T(x|y)$  satisfies equation (\ref{kerneleq}), with $P(x)$ a given probability distribution,
then uniform (unbiased) sampling from  $\{x_0, x_1, \ldots, x_t, \ldots\}$, will yield $x_i$ with probability
$P(x_i).$
}\vskip .3cm

Under these condition the probability $P(x)$ is said to be the equilibrium distribution or the
stationary point for $T(x|y)$.

\subsection{Markov Chain Monte Carlo Integration}
From Theorem (1) and Eq. (\ref{montecarlointegration}) it follows immediately that from any
subsequence of a Markov Chain we can get an unbiased estimation of a function average, that is:
\begin{equation}
	\lim_{T\to\infty} \frac{1}{T}\sum_{t=t_s}^{t_s+T} g(x_t) = E\left[g(x)\right].
\end{equation}
As it turns out, most of the time it is quite impossible to sample directly from a given distribution,
but it is remarkably simple to create a Markov Chain that admits that same distribution as its stationary
point.

\subsection{Detailed Balance}

The transition probability $T(x|y)$ is a real probability in x, that is
\[
	\int dx T(x|y) = 1,
\]
and we can easily see that a sufficient condition for Eq.(\ref{kerneleq}) to hold is to have:
\begin{equation}
	T(x|y) P(y) = T(y|x) P(x).
\label{detailedbalance}
\end{equation}
This equation is called the
detailed balance equation. From detailed balance, equation (\ref{kerneleq} ) follows directly after integrating 
in $y$ both sides of Eq.(\ref{detailedbalance}).

The desired $T(y|x)$ can be built in virtue of the following:

\par{\bf Theorem 2:}{\it
Let $\alpha(y|x)$ be any transition probability, then $P(y)$ will be a stationary distribution for $T(y|x)$ if
\begin{equation}
\label{transitionmatrix}
	T(y|x) = min\bigg(1,\frac{\alpha(x|y)P(y)}{\alpha(y|x)P(x)}\bigg)\alpha(y|x).
\end{equation}
}\vskip .3cm

A particularly simple condition of application of this theorem is when $\alpha(x|y) = \alpha(y|x)$, 
in which case the prescription to build the transition matrix becomes:
\begin{itemize}
\item from a point $x_t$ propose a new point $y$ with probability $\alpha(y|x_t)$;
\item if $P(y) >= P(x_t)$ then set $x_{t+1} = y$, otherwise with probability
 $P(y)/P(x_t)$ set $x_{t+1} = y,$ and with probability $1 - P(y)/P(x_t)$ set 
$ x_{t+1} = x_t.$
\end{itemize}

\section{MCMC for portfolio optimisation}
\label	{integration}
If we choose the transition probability:
\[
\alpha(\vec{p}_{t+1},\vec{m}_{t+1}|\vec{p}_t,\vec{m}_t) = P_{p}(\vec{p}_{t+1}|\vec{m}_{t+1}) P_0(\vec{m}_{t+1})
\]
we have, according to Eq: (\ref{transitionmatrix}),
\[
T(\vec{p}_{t+1},\vec{m}_{t+1}|\vec{p}_t,\vec{m}_t) = \alpha(\vec{p}_{t+1},\vec{m}_{t+1}|\vec{p}_t,\vec{m}_t).
\]
Such a transition is readily obtained by sampling $\vec{m}_{t+1}$ from the distribution 
$P_0(\vec{m})$, then, holding fixed $\vec{m}_{t+1}$, sampling $\vec{p}_{t+1}$ from the 
full conditional $P_{p}(\vec{p}|\vec{m})$.

Sampling from $P_0(\vec{m})$ offers no challenge given that the random variable is normally distributed;
the whole challenge is sampling $\vec{p}$ from its fully conditional distribution.

This can be done by devising a suitable Markov chain with stationary distribution $P(\vec{p}|\vec{m})$.

\subsection{Metropolis MCMC}
The first algorithm we present is a very simple implementation of the evergreen Metropolis
algorithm.

From a location $\vec{p}_t$, generate a random vector $\vec{v}_t$ and consider the point
\begin{equation}
	\vec{q} = \vec{p}_t + \epsilon \vec{v}_t,
\label{step}
\end{equation}

where $\epsilon$ is a small number. 
Let 
\[
	\Delta U  \equiv  E[u(\vec{q}, L, \lambda, T)|\vec{m}] - E[u(\vec{p}, L, \lambda, T)|\vec{m}],
\]
and with probability
\begin{equation}
	\pi  = \mbox{min}\left(1, \exp\bigg( N \Delta U \bigg)\right)
\end{equation}
set $\vec{p}_{t+1} = \vec{q}, $ and with probability $ 1- \pi$, set $ \vec{p}_{t+1} = \vec{p}_t.$

The step described in equation (\ref{step}) guarantees that the transition probability for the process
$ \vec{p}_t \to \vec{p}_{t+1} $ is the same as the transition probability for the inverse process
$ \vec{p}_{t+1} \to \vec{p}_t $.  This suffices to prove that the Markov chain has the desired equilibrium distribution.
The only warning to be issued concerns the range of the variables $\vec{p}.$ The domain $D(\vec{p})$ is bounded
therefore it will happen that step (\ref{step})  will try to get on the outside. In this case care must be taken
to bounce properly (a billiard ball rule will suffice) the trajectory in order to keep the point inside the domain.

\subsection{Hamiltonian MCMC}
The second algorithm we present is well known in the physics literature with the name
of hybrid Monte-Carlo (\cite{Duane87}).
In this appendix we limit ourselves to a short introduction.

Since we have to sample $\vec{p}$ keeping $\vec{m}$ fixed we are only interested in
the functional form on $\vec{p}$ of the full conditional $P_{p}(\vec{p}|\vec{m})$:
\[
	P_{p}(\vec{p}|\vec{m}) \sim exp\bigg( U(\vec{p}) \bigg).
\]
Expectations of functions of $\vec{p}$ will not be affected if we replace $P_{p}(\vec{p}|\vec{m})$
with the distribution
\begin{equation}
	G(\vec{p}, \vec{\pi}|\vec{m}) \sim exp\bigg( U(\vec{p}) - \frac{\vec{\pi}\cdot\vec{\pi}}{2}\bigg),
\end{equation}
then starting from a pair $(\vec{p}_n, \vec{\pi}_n)$, the updating rule is defined as follows:

\begin{itemize}

\item[{\bf Step 1}] Sample $\vec{\eta} $ as a normal variable with mean zero and variance 1.
\item[{\bf Step 2}] For a time interval T, integrate Hamiltons equations 
		\begin{eqnarray}
		\frac{d\pi_i}{dt} & = & -\frac{\partial U}{\partial \pi_i} \\
		 \frac{dp_i}{dt} & = & \pi_i,
		\end{eqnarray}
toghether with the boundary conditions
		\begin{eqnarray}
			\vec{\pi}(0) & = & \vec{\eta}, \\
                        \vec{p}(0) & = & \vec{p}_n;
		\end{eqnarray}
\item[{\bf Step 3}]  With probability
\begin{equation}
	\beta = \mbox{min}\bigg(1, \exp(G(\vec{p}(T),\pi(T)) - G(\vec{p}_n, \eta) \bigg),
\end{equation}
set $\vec{p}_{n+1} = \vec{p}(T)$, and with probability $1 - \beta$ set $\vec{p}_{n+1} = \vec{p}_n$.
\end{itemize}

The clever idea behind this algorithm  rests on the observation that, if step 2 is carried out exactly,
Hamilton's equation enforce $ G(\vec{p}(T),\pi(T)) =  G(\vec{p}_n, \eta) $ and every proposed
configuration is accepted. In general the acceptance rate will be controlled by the numerical accuracy of our
integration scheme. A good scheme is the interleaved leap frog that, for 
finite integration step $\Delta t$ and fixed trajectory length
(that is, scaling the number of steps in the integration scheme with $1/\Delta t$), is guaranteed to have errors $O(\Delta t^2) $.

\markboth{ba\_priors.tex}{ba\_priors.tex}

\section{Likelihoods, Priors and Posteriors}
\label{priors}

\subsection{Likelihoods of Data}
The likelihood of observed data or the conditional density of data w.r.t a given
model $\{\vec{m}, \vec\Omega\}$ is given by:
\begin{eqnarray}
\fl	L(\{R\}|\vec{m}, \vec\Omega) & = &
	  \prod_{n=1}^N \exp\left(
         -\frac{(\vec{r}_n - \vec{m})^T\vec\Omega^{-1}(\vec{r}_n -\vec{m})}{2}
           \right)
	   \frac{d\nu(\vec{r}_n)}{\sqrt{(2\pi)^J|\vec\Omega|}} \\
 & = &  \exp\left( -\frac{1}{2}\sum_{n=1}^N (\vec{r}_n - \vec{m})^T\vec\Omega^{-1}(\vec{r}_n -\vec{m})
           \right)
	   \prod_{n=1}^N\frac{d\nu(\vec{r}_n)}{\sqrt{(2\pi)^J|\vec\Omega|}} \nonumber \\
 & = &  \exp\left( 
  -\frac{N}{2}\vec{m}^T\vec\Omega^{-1}\vec{m} + N\vec{m}\vec\Omega^{-1}\vec{\overline{r}} 
       - \frac{1}{2}\sum_{n=1}^N \vec{r}_n^T\vec\Omega^{-1}\vec{r}_n )
           \right)
	   \prod_{n=1}^N\frac{d\nu(\vec{r}_n)}{\sqrt{(2\pi)^J|\vec\Omega|}}  \nonumber
\end{eqnarray}
where 
\[ \vec{\overline{r}} = (1/N)\sum_{n=1}^N\vec{r}_n\].

The exponent can be written as:
\begin{eqnarray}
  -\frac{N}{2}(\vec{m} - \vec{\overline{r}})^T\vec\Omega^{-1}(\vec{m} - \vec{\overline{r}})
  +\frac{N}{2}\vec{\overline{r}}^T\vec\Omega^{-1}\vec{\overline{r}}
  - \frac{1}{2}\sum_{n=1}^N \vec{r}_n^T\vec\Omega^{-1}\vec{r}_n	  & = &  \nonumber\\                                                                       
  -\frac{N}{2}(\vec{m} - \vec{\overline{r}})^T\vec\Omega^{-1}(\vec{m} - \vec{\overline{r}}) 
  - \frac{1}{2}\sum_{n=1}^N (\vec{r}_n -\vec{\overline{r}})^T\vec\Omega^{-1}(\vec{r}_n - \vec{\overline{r}}) & = &  \\
  -\frac{N}{2}(\vec{m} - \vec{\overline{r}})^T\vec\Omega^{-1}(\vec{m} - \vec{\overline{r}}) 
  - \frac{N}{2} Tr\left[\vec\Omega^{-1}\Sigma\right] &  & , \nonumber
\end{eqnarray}
where $\Sigma$ is the symmetric matrix whose element $\{ij\}$ is:
\[
	\Sigma_{ij} = \frac{1}{N}\sum_{n=1}^N (\vec{r}_n-\vec{\overline{r}})_i
	                                 (\vec{r}_n-\vec{\overline{r}})_j
\]

\subsection{Priors}

The prior for the model $\{\vec{m}, \vec\Omega\}$ is given by:

\begin{equation}
    \Pi_0(\vec{m}, \vec\Omega | I) = \Pi_{m}(\vec{m}|\vec\Omega, I) \Pi_{\vec\Omega}(\vec\Omega | I)
\end{equation}
where:
\[ \Pi_{m}(\vec{m}| I) = 
	  \exp\left(
         -\frac{\beta}{2}(\vec{m} - \vec{\chi})^T\vec\Omega^{-1}(\vec{m} -\vec{\chi})
           \right)
	   \frac{d\nu(\vec{m})}{\sqrt{(2\pi)^J|\vec\Omega|}},
\]

and
\[
  \Pi_{\vec\Omega}(\vec\Omega|I) = K |\vec\Omega|^{-\frac{h+J+1}{2}} \exp\bigg(
  -\frac{h}{2}
   Tr\left[\vec{\Omega^{-1}}\Sigma_{\vec{\Omega}}\right]
  \bigg)d\mu(\vec\Omega),
\]
witk $K$ a constant of proportionality independent of $\vec\Omega$.

The measure $d\nu(\vec{x})$ is a measure in $R^J$  while $d\mu(\vec\Omega)$ is a measure in the
space of symmetric positive definite matrices.
Since any symmetric positive definite matrix $\vec\Omega$ admits a unique decomposition
\[
\vec\Omega = O^T\Lambda O,
\]
where $O$ is an ortogonal matrix in J dimensions and
\[ \Lambda_{ij} = \lambda_j\delta_{ij}, \] 
a diagonal positive definite matrix, the measure $d\mu(\vec\Omega)$ decomposes in
\[ 
  d\mu(\vec\Omega) = d\nu_{+}(\vec{\lambda}) \left[O^TdO\right],
\]
where  $d\nu_{+}(\vec{\lambda})$ is the flat measure in the semisphere $R_{+}^J$ 
and $\left[O^TdO\right] $ is the Haar measure on the orthogonal group in J dimensions.

\subsection{Marginal Distributions}
The marginal distribution for the observed data $M(\{R\}|I)$ is given by:
\begin{eqnarray}
\fl 	\int_{\vec{m},\vec\Omega} L(\{R\}| \vec{m}, \vec\Omega, I) \Pi_0(\vec{m}, \vec\Omega| I)
	   = & & \nonumber \\
   K \prod_{n=1}^N\frac{d\nu(\vec{r}_n)}{\sqrt{(2\pi)^J}} 
  \int_{\vec{m},\vec\Omega} \frac{d\nu(\vec{m})}{\sqrt{(2\pi)^J|\vec\Omega|}}
         d\mu(\vec\Omega)|\vec\Omega|^{-\frac{h+N+J+1}{2}} \exp\bigg(H(\vec{m}, \vec\Omega)\bigg) & &
\end{eqnarray}
where

\begin{eqnarray}
\fl
H(\vec{m}, \vec\Omega) & =  &
  -\frac{N}{2}(\vec{m} - \vec{\overline{r}})^T\vec\Omega^{-1}(\vec{m} - \vec{\overline{r}}) 
         -\frac{\beta}{2}(\vec{m} - \vec{\chi})^T\vec\Omega^{-1}(\vec{m} -\vec{\chi}) \nonumber \\
  & - & \frac{N}{2} Tr\left[\vec\Omega^{-1}\Sigma\right] 
  -\frac{h}{2}Tr\left[\vec\Omega^{-1}\Sigma_{\vec\Omega}\right] \\
   & & \nonumber\\
   & = &   -\frac{N+\beta}{2}\vec{m}^{\ T}\vec\Omega^{-1}\vec{m}
         +N \vec{m}^{\ T}\vec\Omega^{-1}(\vec{\overline{r}}+\kappa\vec{\chi}) \nonumber\\
         & - & \frac{N}{2}\vec{\overline{r}}^{\ T}\vec\Omega^{-1}\vec{\overline{r}}
          -  \frac{\beta}{2}\vec{\chi}^{\ T}\vec\Omega^{-1}\vec{\chi} \nonumber\\
  & - & \frac{N+h}{2}Tr\left[\vec\Omega^{-1}\left( \frac{h\Sigma_{\vec\Omega}}{N+h} +  \frac{N\Sigma}{N+h} \right) \right] \\
  & & \nonumber\\
  & = &   -\frac{N(1+\kappa)}{2}(\vec{m}-\vec{M})^{\ T}\vec\Omega^{-1}(\vec{m} - \vec{M}) \nonumber\\
         & - & \frac{N}{2}Tr\bigg[\vec\Omega^{-1}\bigg(
\vec{\overline{r}} \vec{\overline{r}}^{\ T} - (1+\kappa)\vec{M}\vec{M}^{\ T} 
        + \kappa\vec{\chi}\vec{\chi}^{\ T} \bigg)\bigg] \nonumber\\
  & - & \frac{N+h}{2}Tr\left[\vec\Omega^{-1}\left( \frac{h\Sigma_{\vec\Omega}}{N+h} +  \frac{N\Sigma}{N+h} \right) \right] 
\end{eqnarray}

with:
\begin{eqnarray}
	\kappa & = & \frac{\beta}{N} \\
	\vec{M} & = & \frac{\vec{\overline{r}} + \kappa\vec\chi}{1+\vec{\kappa}}.
\end{eqnarray}

If we define the matrix
\[
	C =  \vec{\overline{r}}\vec{\overline{r}}^{\ T}
		+ \kappa\vec{\chi}\vec{\chi}^{\ T} - (1+\kappa)\vec{M}\vec{M}^{\ T}
\]
we get:
\begin{eqnarray}
	C & = &  \frac{1}{1+\kappa}\left((1+\kappa)\vec{\overline{r}}\vec{\overline{r}}^{\ T}
		+ \kappa (1+\kappa) \vec{\chi}\vec{\chi}^{\ T} 
                - (\vec{\overline{r}}+\kappa\vec\chi)(\vec{\overline{r}}+\kappa\vec\chi)^{\ T}
	\right) \\
	 & = &  \frac{1}{1+\kappa}\left(\kappa\vec{\overline{r}}\vec{\overline{r}}^{\ T}
		+ \kappa \vec{\chi}\vec{\chi}^{\ T} 
                - \kappa\vec{\overline{r}}\vec\chi^{\ T}
                - \kappa\vec\chi\vec{\overline{r}}^{\ T}
        \right) \\
	 & = &  \frac{\kappa}{1+\kappa}(\vec{\overline{r}}-\vec\chi)(\vec{\overline{r}}-\vec\chi)^{\ T}
\end{eqnarray}

by which we get the final expression:

\begin{equation}
\fl
H(\vec{m}, \vec\Omega)  =
   -\frac{N(1+\kappa)}{2}(\vec{m}-\vec{M})^{\ T}\vec\Omega^{-1}(\vec{m} - \vec{M})
   - \frac{N+h}{2}Tr\left[\vec\Omega^{-1} A\right] ,
\end{equation}

where the matrix A is given by:
\begin{equation}
	A   =  \frac{h\Sigma_{\vec\Omega}}{N+h} +  \frac{N\Sigma}{N+h} 
	   +    \frac{\kappa(\vec{\overline{r}}-\vec\chi)(\vec{\overline{r}}-\vec\chi)^{\ T}}{(1+\kappa)(N+h)}.
\end{equation}

The posterior is characterized by the following:
\begin{eqnarray}
	\vec{m}|\vec\Omega & \sim & N(\frac{\vec{\overline{r}} + \kappa\vec\chi}{1+\kappa}, \vec\Omega) \\
        \vec\Omega  & \sim & W^{-1}(N+h, A).
\end{eqnarray}

\section*{References}
\bibliographystyle{plain}

\begin{thebibliography}{1}
\bibitem{Ait}
Ait-Sahalia Y and Brandt M 2001
Variable Selection for Portfolio Choice
{\it Journal of Finance} {\bf 56} 1297-1351

\bibitem {Balduzzi}
 Balduzzi P and Liu L 2000
 Parameter Uncertainty and International Investment
 {\it Working Paper, Boston College}

\bibitem{Barberis}
Barberis N 2000 
Investing for the Long Run when Returns are Predictable
{\it Journal of Finance} {\bf 55} 225-264

\bibitem{Barry}
 Barry CB 1974
 Portfolio Analysis under Uncertain Means, Variances and Covariances
 {\it Journal of Finance} {\bf 29} 515-522

\bibitem{Bawa}
 Bawa V Brown S and Klein R 1979
 Estimation Risk and Optimal Portfolio Choice
 {\it North Holland, Amsterdam}

 \bibitem{BlackLitterman}
 Black F and Litterman R 1992 
 Global Portfolio Optimization
 {\it Financial Analyst Journal} {\bf 09} 28-43

 \bibitem{Brennan}
Brennan M 1998 
The Role of Learning in Dynamic Portfolio Decisions
{\it European Finance Review} {\bf 1} 295-306

\bibitem{Bouchaud} 
Bouchaud JP and Potters M 2000 
Theory of Financial Risks
{\it Cambridge University Press}

\bibitem{Bouchaud2} 
Bouchaud JP, Potters M and Aguilar JP 1997 
Missing Information and Asset Allocation
{\it Unpublished paper: available for download at www.science-finance.fr}

 \bibitem{CampbellViceira01}
 Campbell J and Viceira LM 2001
 Who Should Buy Long-Term Bonds?
 {\it American Economic Review} {\bf 91} 99-127
 
 \bibitem{CampbellViceira02}
 Campbell J and Viceira LM 2002 
 Strategic Asset Allocation: Portfolio Choice for Long-Term Investors
 {\it Clarendon Lectures in Economics, Oxford University Press}

 \bibitem{Ziemba}
 Chopra V and Zembia WT 1993
 The Effect of Errors in Means, Variances and Covariances on Optimal Portfolio Choice
{\it Journal of Portfolio Management} {\bf Winter 1993} 6-11

 \bibitem{Consiglio-et-all01} Consiglio A, Cocco F and Zenios S A 2001
 The value of integrative risk management for insurance products with guarantees
{\it Journal of Risk Finance} {\bf 2} 6-16

\bibitem{Duane87} Duane S, Kennedy AD, Pendleton BJ and Roweth D 1987 
Hybrid Monte Carlo
{\it Phys.Lett.} {\bf B195} 216-222

\bibitem{Frost}
 Frost P and Savarino J 1986
 An Empirical Bayes Approach to Efficient Portfolio Selection
 {\it Journal of Financial and Quantitative Analysis} {\bf 21} 293-305

\bibitem{Gilks96} 
Gilks WR, Richardson S and Spiegelhalter D J 1996
Markov Chain Monte Carlo in Practice
{\it Chapman \& Hall}

\bibitem{Hirsh} 
Hirshleifer D
Investor psychology and asset pricing 2001
{\it Journal of Finance} {\bf 56} 1533-1597 

\bibitem{TerHorst}
ter Horst JR, de Roon FA and Weker BJM 2002
Incorporating Estimation Risk in Portfolio Choice
{\it Working Paper, Tilburg University}

\bibitem{Ingersoll} Ingersoll J 1987
Theory of Financial Decision Making
{\it Rowman \& Littlefield }

\bibitem{Kirkpatrick83}
Kirkpatrick S, Gelatt CD and Vecchi MP 1983
Optimisation by simulated annealing
{\it Science} {\bf 220} (4598), 671-680

\bibitem{KritzmanRich}
Kritzman M and Rich D 1998 
Beware of the Dogma: The truth about Time Diversification
{\it The Journal of Portfolio Management}, {\bf 24} 66-77

\bibitem{Jobson}
 Jobson JD and Korkie R 1980
 Estimation for Markowitz Efficient Portfolios
 {\it Journal of the American Statistical Association} {\bf 75} 544-554

\bibitem{Polson02}
Johannes M and Polson N 2002 
MCMC methods in Financial Econometrics 
{\it Handbook of Financial Econometrics} Forthcoming 

\bibitem{Johannes}							  
 Johannes M, Polson N and Stroud J  2002
 Sequential Optimal Portfolio Performance: Market and Volatility Timing
 {\it Working Paper, Columbia University} 

\bibitem{Jorion}
Jorion Ph 1985
International Portfolio Diversification with Estimation Risk
{\it Journal of Business} {\bf 58} 259-278

\bibitem{Markovitz}
Markowitz H 1952
Portfolio Selection
{\it Journal of Finance} {\bf 7} 77-91

\bibitem{Michaud}
Michaud RO 1989
The Markowitz Optimization Enigma: Is Optimized Optimal?
{\it Financial Analyst Journal} {\bf 45} 31-42

\bibitem{Samuelson94}
Samuelson PA 1994
The Long-Term Case for Equities
{\it The Journal of Portfolio Managment} {\bf 21}

\bibitem{SThorley}
Thorley S 1995
The Time Diversification Controversy
{\it Financial Analyst Journal} {\bf 06} 68-76

\end{thebibliography}

\end{document}